\documentclass[structabstract]{aa}  
\usepackage{graphicx}
\usepackage{natbib}
\bibpunct{(}{)}{;}{a}{}{,}
\newcommand{\lesssim}{\mathrel{\hbox{\rlap{\hbox{\lower4pt\hbox{$\sim$}}}\hbox{$<$}}}}
\newcommand{\gtrsim}{\mathrel{\hbox{\rlap{\hbox{\lower4pt\hbox{$\sim$}}}\hbox{$>$}}}}
\usepackage{txfonts}
\begin{document}
   \title{The circumstellar environment
of the YSO TMR-1 and a revisit to the candidate very low-mass
object TMR-1C \thanks{Based
on observations made with ESO telescopes at the La Silla Paranal Observatory
under programme ID 265.C-5747(A), and based on data obtained from the ESO
Science Archive Facility}}

  \author{Monika G.\ Petr-Gotzens \inst{1}, Jean-Gabriel Cuby \inst{2},
           Michael D.\ Smith \inst{3}, \and Michael F.\ Sterzik \inst{4}
          }

   \institute{European Southern Observatory, Karl-Schwarzschild-Str.\ 2,
              D-85748 Garching, Germany \and
             Laboratoire d'Astrophysique de Marseille, OAMP, Universit\'e 
Aix-Marseille \& CNRS, 38 rue Fr\'ed\'eric Joliot Curie, 13388 Marseille 
cedex 13, France
              \and
             Centre for Astrophysics and Planetary Science,
         University of Kent, Canterbury, Kent CT2 7NR, England
             \and
             European Southern Observatory, Casilla 19001,
              Santiago 19, Chile
             }
\authorrunning{Petr-Gotzens et al.}
\titlerunning{The circumstellar environment
of the YSO TMR-1 and the candidate very low-mass object TMR-1C}

   \date{Received 26 May 2009; accepted 27 July, 2010}

 
  \abstract
   {TMR-1 (IRAS~04361+2547) is a class~I proto-stellar source located in the nearby Taurus star-forming region. Its circumstellar
    environment is characterized by extended dust emission with complex structures and conspicuous filaments.
    A faint companion, called TMR-1C,  located near the proto-star had been detected in previous studies, but its nature as a 
    very young substellar object remained inconclusive. 
   }
   {We aim at improving the constraints on the nature of the faint object TMR-1C, and to investigate 
   the process of very low-mass star formation in the TMR-1 system. }
   {Using very sensitive infrared imaging observations of the TMR-1 system as well as near-infrared spectroscopy, we
   construct the spectral energy distribution of TMR-1C over a much larger wavelength range as had been possible
   in previous work. We then compare the  spectral energy distribution with models of extincted background stars,
  young sub-stellar objects, and very low-mass stars with circumstellar disk and envelope emission.  We also
  search for additional low-luminosity sources in the immediate environment of the TMR-1 proto-stellar
  source.
   Furthermore, we study the surrounding near-infrared dust morphology,
   and analyse the emission line spectrum of a filamentary structure in the physical context of a bow-shock model.
   }
   {We find that the observed spectral energy distribution of TMR-1C is inconsistent with an extincted background star, 
   nor can be fitted with available models for a young extremely low-mass ($\lesssim 12{\rm M}_{\rm J}$) object. 
   Our near-infrared spectrum indicates an effective temperature $\gtrsim$3000K. Based on a good match of 
   TMR-1C's spectral energy distribution with radiation transfer models of young stellar objects with circumstellar
   disks, we propose
   that TMR-1C is most likely a very low-mass star with M$\approx 0.1-0.2{\rm M}_{\odot}$ surrounded by a circumstellar
    disk with high inclination, $i>80^{\circ}$. Interestingly, we detect an additional very faint source, which we call  
    TMR-1D, and that shows a quite striking symmetry in position with TMR-1C. TMR-1C and TMR-1D may have been
    formed from a common triggered star-formation event, caused by a powerful outflow or by the collision of 
    primordial proto-stellar disks. The impact of an outflow is traced by molecular hydrogen emission that we detect from 
    a distinct filament pointing towards TMR-1C. A comparison with C-type bow shock models confirms that the 
    emission is caused by shock excitation.}
   {}


   \maketitle
%

\section{Introduction}
TMR-1 (IRAS~04361+2547) is a class I young stellar object located in the Taurus molecular cloud,
that was actually resolved into a binary source with a measured components' separation of 0.31\arcsec 
(Terebey et al. 1998, hereafter T98). We will therefore refer to it as TMR-1AB in the following. 
The total bolometric luminosity of TMR-1AB had been estimated to $\sim 2.8{\rm L}_\odot$, indicative for a 
low-mass protostellar system (Kenyon et al.\  1993). The circumstellar environment of TMR-1AB
is characterized by 
extended emission from a dusty proto-stellar envelope and from patches of
molecular cloud material left over from the proto-stellar collapse.    
In their sensitive HST/NICMOS observations, T98 detected a
faint compact object, named TMR-1C, at a projected distance of $\sim$ 10\arcsec \ from TMR-1AB, which 
corresponds to $\sim$1400\,AU at the distance of the Taurus molecular cloud.
The physical association of TMR-1C with TMR-1AB was suggested on the basis of the presence of
a striking filament structure that arises from TMR-1 and points directly
towards TMR-1C. T98 further suggested that TMR1-C was catapulted to its current location 
due to dynamical interactions with the proto-binary TMR1-AB 
and that the arc-shaped filament could trace the ejection
path of TMR-1C through the gaseous infalling circumstellar envelope of TMR-1. 
The very low luminosity suggested for TMR-1C indicated that it should be a substellar
object, maybe even a planetary mass object.
However, the physical association
of TMR-1C with TMR-1, and hence its nature as a substellar object, was strongly debated
during the years after its discovery. In an attempt to clarify the evolutionary status of TMR-1C,
Terebey et al.\ (2000) carried out near-infrared spectroscopy using the Keck telescope.
The result of these observations showed that the spectrum of TMR-1C, at the signal-to-noise 
level that could be reached, is consistent with an extincted background dwarf star spectrum,
but still room was left for an interpretation within the
extremely low-mass object ejection hypothesis.

In this paper we use ESO {\it Very Large Telescope} (VLT) data
obtained with the {\it Infrared Spectrometer and Array Camera} (ISAAC),
as well as {\it Spitzer}/IRAC observations in order to revisit the TMR-1 system.
After the description of the observations and the data reduction (Sec.\ 2), we discuss  
in Sec.\  3.1 the morphology of the circumstellar dusty environment of TMR-1
and report on the detection of new features and objects identified in our
sensitive ISAAC images.
We then use our K-band low-resolution spectroscopy together with the  
spectral energy distribution of TMR-1C, which we construct from the photometry presented in
this paper and collected from the literature, in order to analyze the nature of TMR-1C
in Sec.\ 3.2 and Sec.\  3.3. Our K-band spectroscopy was also performed on a 
significant part of the filament  structure ''connecting'' TMR-1C with TMR-1AB. Since
the spectral resolution is almost 4 times higher than in previous spectroscopic observations 
we are able to resolve numerous emission-line features arising from the filament. In 
Sec.\  3.4 we analyze the filament spectrum in detail, which leads us to conclude that
a significant part of its emission is characterized by shock-excited emission. In Sec.\ 4 we present 
scenarios that could possibly explain the symmetry of the circumstellar nebulosities,
the filamentary structures, and the existence of a pair of very low-mass objects, as 
a physical entity. Finally, we summarize our results and conclusions in Sec.\ 5.

   \begin{figure*}
      \resizebox{17cm}{!}{\includegraphics{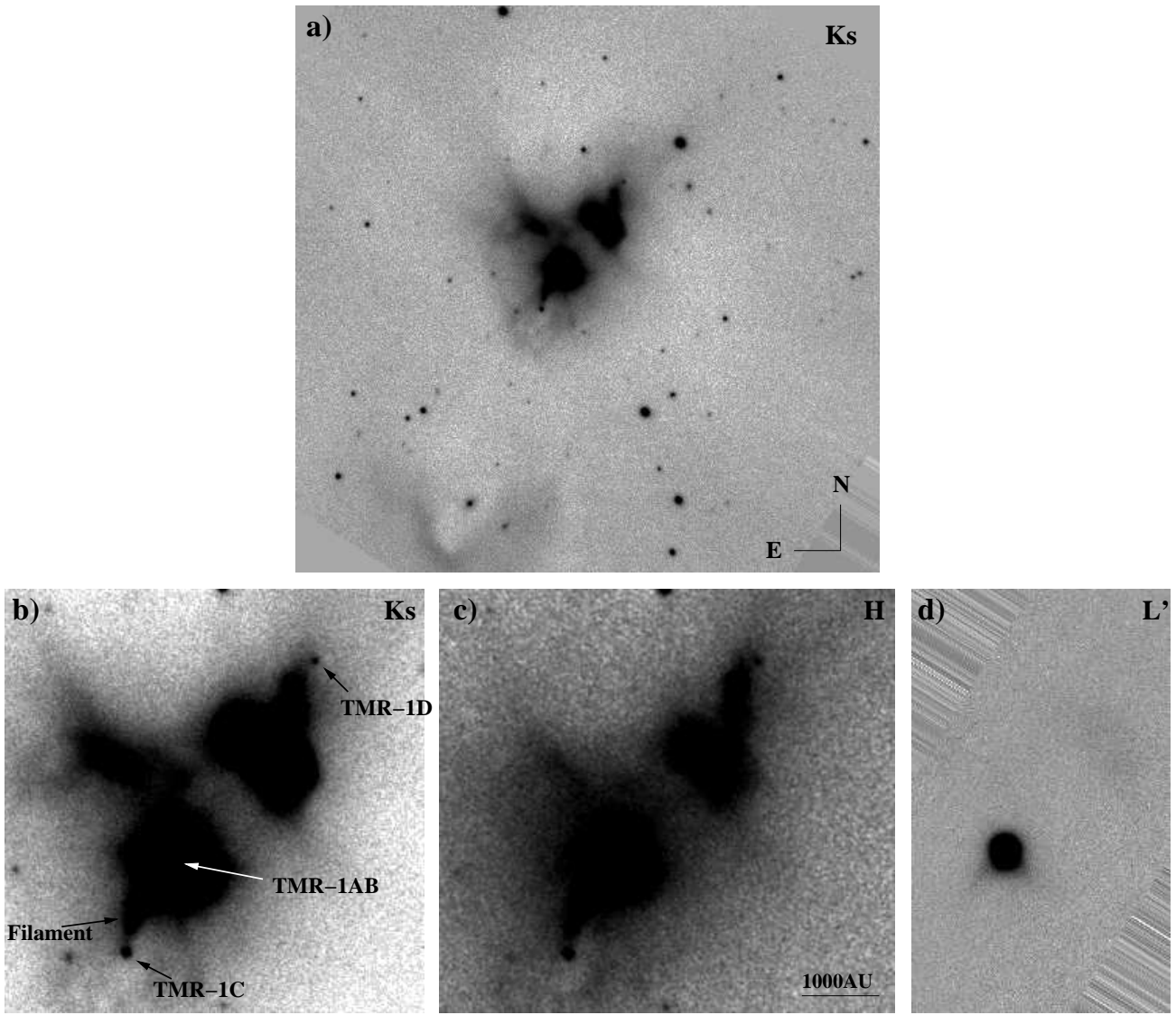}}
      \hfill
      \caption[ ] { {\bf a)} ISAAC Ks (2.16$\mu$m) image of TMR-1 and its surrounding.      
       The total integration time in this image is 20 minutes and the image size is approximately  
       $2.4^{\prime}\times2.4^{\prime}$  
       {\bf b)} Zoom-in on the Ks-image of TMR-1 clearly showing the candidate very low-mass object 
       TMR-1C at the end of the filament structure arising from the nebulosity around the  binary 
       TMR-1AB, and our detection of a 'counter-object', TMR-1D. Image size is approximately  
       $40^{\prime\prime}\times40^{\prime\prime}$ {\bf c)} H-band image of the same field 
       as shown in b). {\bf d)} L$^{\prime}$-band image of TMR-1, at the same scale and orientation as the fields shown
       in b) and c).}
      \label{K_full_field}
    \end{figure*}


\section{Imaging and Spectroscopy Data}

\subsection{Near-infrared photometry from VLT/ISAAC HK$_s$ observations}
\label{photom}
A first set of images of TMR-1 was obtained at Ks-band ($\lambda_c=2.16\mu$m, $\Delta \lambda=0.27\mu$m)
with ISAAC at the VLT-ANTU(UT1) telescope as part of commissioning of
the instrument. These data were taken during the night December 04-05, 1998,  immediately 
before the spectroscopic observations, providing also a flux 
calibration for these (Sec.~\ref{Obs_Spec}). The data are accessible via the ESO Science Archive
Facility\footnote{http://archive.eso.org/eso/eso\_archive\_main.html}.
The scale of the ISAAC/Hawaii Rockwell detector was
0.147\arcsec/pix, yielding a field-of-view of 2.5\arcmin$\times$2.5\arcmin.
The images typically consisted of 50 frames averaged with 1.77
seconds individual exposure time. Those having the best image quality, of the order 
of $\sim 0.55^{\prime\prime}$ FWHM, were
selected for further data reduction. The sky contribution was determined
from median combining the individual images, and then subtracted from each image. 
Finally, the sky subtracted images were
badpixel corrected, aligned and combined, and trimmed to a common overlap 
region. The total integration time in the final mosaicked image was 177\,sec. 

  \begin{table*}
      \caption[]{Infrared photometry of TMR-1C}
         \label{phot}
      \begin{tabular}{clcc}
            \hline
            \noalign{\smallskip}
             Filter & Central& Magnitude & Date of observation  \\
                    & Wavelength   & or Flux &   (UT) \\
                    & ($\mu$m) &  (mag or Jy) & \\
            \noalign{\smallskip}
            \hline
            \noalign{\smallskip}
H      & 1.65 & 19.21$\pm$0.05 & Oct 01 + 02, 2000 \\
K$_s$  & 2.16 & 17.63$\pm$0.1 & Dec 05, 1998 \\
K$_s$  & 2.16 & 17.53$\pm$0.05 & Oct 01, 2000 \\
L$^{\prime}$ & 3.78 & $<7.5\times10^{-5}$ & Sept 20 + Oct 02, 2000  \\
IRAC CH1 & 3.6 & $<0.12\times10^{-3}$ & Mar 07, 2004 \\
IRAC CH2 & 4.5 & $<0.51\times10^{-3}$ & Mar 07, 2004 \\
            \noalign{\smallskip}
            \hline
         \end{tabular}
\end{table*}

A second set of ISAAC near-infrared data was obtained at H-band
($\lambda_c=1.65\mu$m, $\Delta \lambda=0.30\mu$m), and again
at K$_S$-band, on two nights during October 2000
(Table~\ref{phot}).  The images  were taken by
using a random jitter pattern centered on TMR-1.
In order to avoid strong saturation
from the brightest source, TMR-1\,AB, short detector integration times of
DIT=2sec and DIT=3sec were used for K$_S$ and H exposures respectively.
The total exposure times per image were 2$\times$20(DIT$\times$NDIT)sec
for K$_S$ images and 3$\times$20(DIT$\times$NDIT)sec for H images.
The data were then processed with the
{\it Eclipse}\footnote{http://www.eso.org/projects/aot/eclipse/} data reduction
package (Devillard 1997), using the {\it jitter} algorithm. 
The images were dark-subtracted and flat-fielded with a skyflat obtained
from exposures taken during twilight, then sky-subtracted, and finally recentered
and stacked to single deep H- and Ks-band images. The total
integration time in the stacked H-image is 12\,minutes, while it is 20minutes
for the Ks-image. TMR-1 and its immediate
surrounding as seen in these sensitive images is shown in Figures~\ref{K_full_field}a, b, and c.
The measured image qualities are very good, with
FWHM$\sim 0.5^{\prime\prime}$ (Ks) and FWHM$\sim 0.7^{\prime\prime}$ (H),
but still not good enough to resolve the binary TMR-1AB.

Accurate photometry for all sources with a minimum S/N detection of 8
was then performed on the H and Ks mosaic images by using 
PSF-fitting with IRAF/{\it daophot}.
In particular for TMR-1C, which is close to
the nebulous filament and the general nebulosity surrounding
TMR1, PSF photometry provides a much more accurate photometric measurement
than aperture photometry. In all of our final Ks and H mosaics 
TMR-1C has been well detected at a signal-to-noise ratio of $\geq$15.
For the observations carried out in 2000, the absolute photometric calibration has been derived
from the UKIRT faint standard FS12 that was observed immediately after our
observations of TMR-1. The resulting photometry for TMR-1C is Ks=17.53\,mag
and H=19.21\,mag with an uncertainty of $\sim 5\%$. We checked the photometric
calibration against the photometry of a 2MASS star that is present in the ISAAC
field (2MASS 04391199+2553490), and find agreement within 0.06\,mag. Our Ks
photometry of TMR-1C is also consistent with values reported by other authors at similar wavelength 
(Itoh et al.\ 1999), and also in agreement with the Ks-photometry determined from
the 1998 imaging data (see Table~\ref{phot}). The photometric accuracy for the 1998 
data is worse due to the lack of standard star observations during the observing night, so that
absolute photometry was derived assuming the Ks-zeropoint determined
for the ISAAC commissioning\,I period. 

Our deep imaging photometry revealed numerous faint objects,  detected all over the
field-of-view (Figure~\ref{K_full_field}a). 
There are more than 20 objects with a Ks-magnitude similar or fainter than
TMR-1C, indicating that we are beginning to penetrate through the large
amount of dust of the molecular cloud core.
In this respect, the faintness and appearance of TMR-1C
seems not to be any special, except for its location at the end of the filament.
Very interestingly, we also detected a previously unknown, faint object which looks like a 
TMR-1C counterpart: it is located close to the
end of a broad filamentary structure that extends from the large nebulosity seen to the North-West
of TMR-1AB (Figure~\ref{K_full_field}a, b and c). 
The object is clearly a $>10\sigma$ detection in the
deep H and Ks-band images and fainter than TMR-1C  by $\sim 1.1$mag at Ks and
by $\sim1.0$mag at H.
We will call this object TMR-1D hereafter. Although at this point we don't have 
a proof that there is any physical relation between this object and TMR-1 or to TMR-1C,
we find it quite striking to detect such a symmetry.
In the course of this paper, we will discuss the presence of TMR-1D in the context of different
scenarios.

   \begin{figure*}
     \resizebox{17cm}{!}{\includegraphics{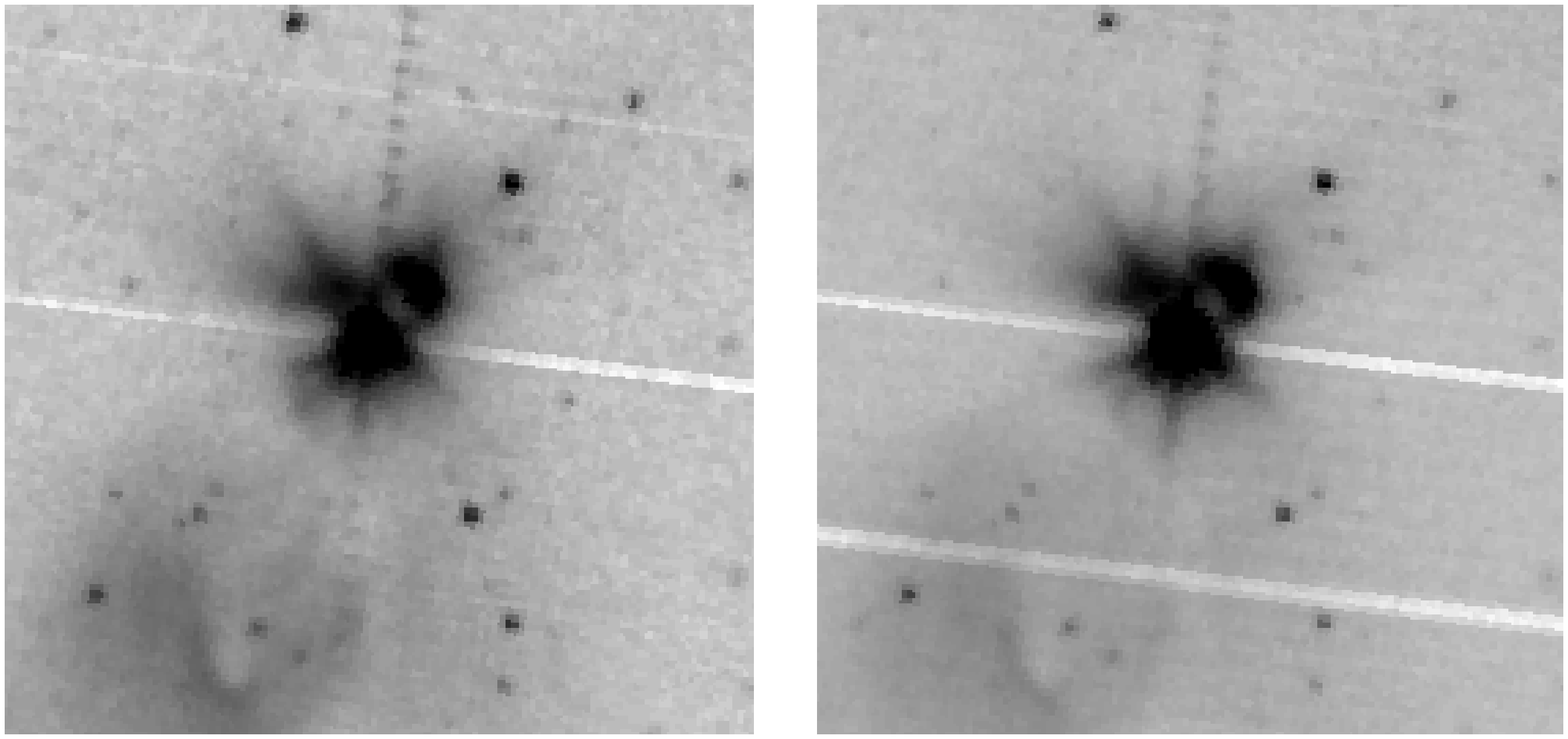}}
      \hfill
      \caption[ ] { {\it Spitzer}/IRAC image at 3.6$\mu$m (left) and 4.5$\mu$m (right) showing TMR-1 and its surrounding. 
      North is up and East is to the left, and the image sizes are the same as in Figure~\ref{K_full_field}a).   
      }
      \label{irac3_6and4_5}
    \end{figure*}

\subsection{Mid-infrared photometry from VLT/ISAAC L$^{\prime}$-band and Spitzer/IRAC observations}
ISAAC mid-infrared imaging at L$^{\prime}$-band
($\lambda_c=3.78\mu$m, $\Delta \lambda=0.58\mu$m), had been carried out with the
long-wavelength imaging camera of ISAAC, equipped with an Aladdin 1024$\times$1024
pix array, and providing a pixel scale of 0.071\arcsec/pix in LWI4 mode.
Data were taken in uncorrelated/high bias read-out mode using standard
chopping and nodding techniques, with a chop-throw position angle of
122$^{\circ}$, an amplitude of 30\arcsec, and a chopping frequency of
0.1~Hz. The reduction process included bad pixel filtering and flat-fielding.
Then, positive and negative beams from the nod A and B positions were
combined. The total integration time yielded for the final image, shown in Figure~1d, is $\sim 3000$sec.
TMR-1C is not detected down to the limiting sensitivity level (3$\sigma$) of
$\sim 7.5\times10^{-5}$\,Jy. Actually, the only source clearly detected in
our L$^{\prime}$-band
image is TMR-1AB itself. There is also a slight hint of nebulous emission
close to TMR-1AB and in the direction of the filament, as well as
a faint detection of diffuse emission from the north-western nebulosity
structure. The object TMR-1D was not detected.

We have also searched the Spitzer archive for data of TMR-1, and found
that observations have been obtained as part of the Spitzer/IRAC programmes
No.\ 37 (PI: Fazio) and No.\ 3584 (PI: Padgett) devoted to IRAC mappings
of the Taurus molecular cloud.
Since the spatial resolution of IRAC at its 4 wavelengths channels, ch1 (3.6mic),
ch2 (4.5mic), ch3 (5.8mic), ch4 (8.0mic) is on the order of 1\arcsec to 2\arcsec and TMR-1C
is almost 10\arcsec away from TMR-1AB, TMR-1C could have been detected if it is bright
enough at IRAC wavelengths. Therefore, we retrieved the post-BCD data from the Spitzer archive,
reduced with the Spitzer pipeline version S14.0, and
investigated the images. 
The diffuse nebulosities of TMR-1 are well detected in the IRAC 3.6mic and 4.5mic images, which
we show in Figure~\ref{irac3_6and4_5}, and they are very similar in 
appearance as in our ISAAC 2.2mic image, despite at much lower spatial resolution. 
The IRAC channel~1 has a central wavelength close to the ISAAC L$^{\prime}$ filter, but due to its
much larger pixels IRAC is more sensitive to extended emission than ISAAC.  At IRAC
channel~3 and 4 (5.8mic and 8.0mic) only the main source TMR-1AB is seen.
On none of the mosaiced IRAC images TMR-1C is detected. This may
partly be due to the closeness of TMR-1C to the very bright main source TMR-1AB, which has
a flux of $\sim$110\,mJy
at 3.6mic, and $\sim$145\,mJy at 4.5mic, but may also be related to an intrinsic
faintness of TMR-1C in this wavelength range. Given the local background and the Poisson 
noise measurement at the
expected position of TMR-1C, we estimate upper 3$\sigma$ flux detection limits of 0.15mJy
(3.6$\mu$m) and 0.51mJy (4.5$\mu$m). Concerning
the 5.8$\mu$m and 8.0$\mu$m images, the PSF wings of the extremely
bright source TMR-1AB prevent any meaningful determination of an upper flux limit.
The upper flux limits extracted from the Spitzer data are listed together with the ISAAC
photometry in Table~\ref{phot}.

\subsection{VLT/ISAAC K-band Spectroscopy}
\label{Obs_Spec}
\begin{figure}
      \resizebox{\hsize}{!}{\includegraphics{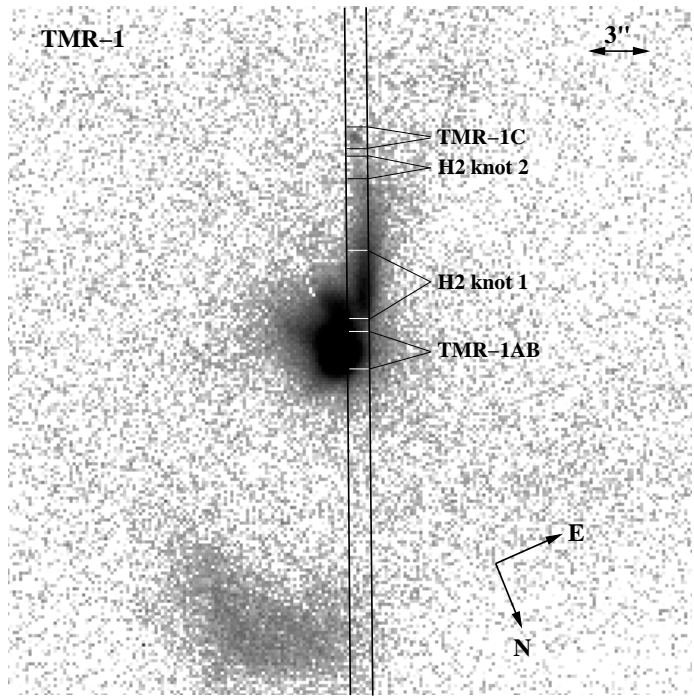}}
      \caption[]{Position of the slit on TMR-1C and the filament
       is shown. The sizes and positions of the apertures for the extracted spectra
      are indicated. The underlying image is the ISAAC Ks aquisition image
       on logarithmic scale.} 
         \label{slitpos}
   \end{figure}

K-band low-resolution (R=450) long-slit spectroscopy of TMR-1C,
and of the filament was performed with ISAAC during the night Dec 04-05, 1998.
In Figure~\ref{slitpos} we show the slit
positioning on TMR-1C and the filament. The width of the slit was 1\arcsec
and spectra were obtained by observing the object alternately at two
positions along the slit, i.e.\ by nodding along the slit with a nodthrow
size of 40\arcsec. An exposure time of 120\,seconds per
individual frame was used and the resulting effective exposure time of the final 
co-added spectrum is 7680\,seconds.
Since the slit was positioned in a way
to include TMR-1C and most of the filament, the binary TMR-1AB
was correspondingly slightly out of the slit and only light from its PSF wings
and from the immediate circumbinary environment entered the slit.

%

The spectral images were sky-subtracted in pairs using the respectively
nodded companion image for subtraction, cleaned of pixels affected by
cosmic rays and non-linear response, and then combined to a 
single spectral image of TMR-1, which is shown in Cuby et al.\ (2000).
Four different apertures have been extracted from this 2D-spectrum
at positions that are shown in Figure~\ref{slitpos}.
The largest aperture (H$_2$~knot~1) actually includes two separate peaks
of molecular hydrogen emission (cf. Figure~\ref{h2exc}). Both knots appeared
to be very similar on the 2D spectral image and a single aperture was therefore
used to extract the spectrum, improving the
signal to noise ratio.

   \begin{table*}
      \caption[]{Observed line fluxes in TMR-1 spectra, normalized to the H$_2$ 1-0 S(1) flux.}
         \label{lines}
      \begin{tabular}{cccc}
            \hline
            \noalign{\smallskip}
            & Spectral & Measured   & Normalized  \\
            &  line & wavelength & integrated line flux  \\
            &       & (\AA)      &  \\
            \noalign{\smallskip}
            \hline
            \noalign{\smallskip}
            & H$_2$ 1-0 S(1) & 21222 & 1.00 $\pm 0.02$  \\
            & Br$\gamma$ (7-4) & 21664 &   \\
  close to   & H$_2$ 1-0 S(0) & 22231 & 0.22  $\pm 0.03$  \\
TMR-1AB & H$_2$ 1-0 Q(1) & 24057 & 1.29 $\pm 0.29$  \\
            & H$_2$ 1-0 Q(3) & 24226 & 0.86 $\pm 0.11$   \\
            & H$_2$ 1-0 Q(4) & 24377 & 0.33 $\pm 0.06$   \\
            & H$_2$ 1-0 Q(5) & 24536 & 0.69 $\pm 0.09$  \\
            \noalign{\smallskip}
            \hline
            \noalign{\smallskip}
            & H$_2$ 1-0 S(3) & 19573 & 0.49 $\pm 0.01$  \\
            & H$_2$ 1-0 S(2) & 20332 & 0.34 $\pm 0.02$ \\
            & H$_2$ 1-0 S(1) & 21211 & 1.00 $\pm 0.004$  \\
            & Br$\gamma$ (7-4) & 21659 &  \\
           & H$_2$ 1-0 S(0) & 22220 & 0.23 $\pm 0.01$  \\
 H$_2$~knot~1 & H$_2$ 2-1 S(1) & 22466 & 0.12 $\pm 0.002$  \\
           & H$_2$ 1-0 Q(1) & 24044 &1.06 $\pm 0.02$  \\
            & H$_2$ 1-0 Q(2) & 24110 & 0.29 $\pm 0.01$  \\
            & H$_2$ 1-0 Q(3) & 24214 & 0.87 $\pm 0.01$  \\
            & H$_2$ 1-0 Q(4) & 24353 & 0.46 $\pm 0.02$ \\
            & H$_2$ 1-0 Q(5) & 24524 & 0.80 $\pm 0.01$ \\
            & H$_2$ 1-0 Q(6) & 24724 & 0.40 $\pm 0.01$  \\
            \noalign{\smallskip}
            \hline
            \noalign{\smallskip}
            & H$_2$ 1-0 S(3) & 19571 & 0.71 $\pm 0.03$  \\
            & H$_2$ 1-0 S(2) & 20328 & 0.26 $\pm 0.05$  \\
            & H$_2$ 1-0 S(1) & 21206 & 1.00 $\pm 0.04$  \\
            & H$_2$ 1-0 S(0) & 22221 & 0.20 $\pm 0.01$  \\
H$_2$~knot~2  & H$_2$ 2-1 S(1) & 22465 & 0.18 $\pm 0.10$ \\
            & H$_2$ 1-0 Q(1) & 24039 & 1.22 $\pm 0.02$  \\
            & H$_2$ 1-0 Q(2) & 24099 & 0.30 $\pm 0.07$  \\
            & H$_2$ 1-0 Q(3) & 24208 & 1.14 $\pm 0.23$  \\
            & H$_2$ 1-0 Q(4) & 24351 & 0.58 $\pm 0.13$ \\
            & H$_2$ 1-0 Q(5) & 24522 & 0.85 $\pm 0.11$ \\
            \noalign{\smallskip}
            \hline
         \end{tabular}
\begin{list}{}{}
\item[] Note: No line fluxes are given for the Br$\gamma$ line,
             because the observed fluxes are overestimated (see main text for further discussion).
\end{list}
\end{table*}

   \begin{figure*}
      \resizebox{14cm}{!}{\includegraphics{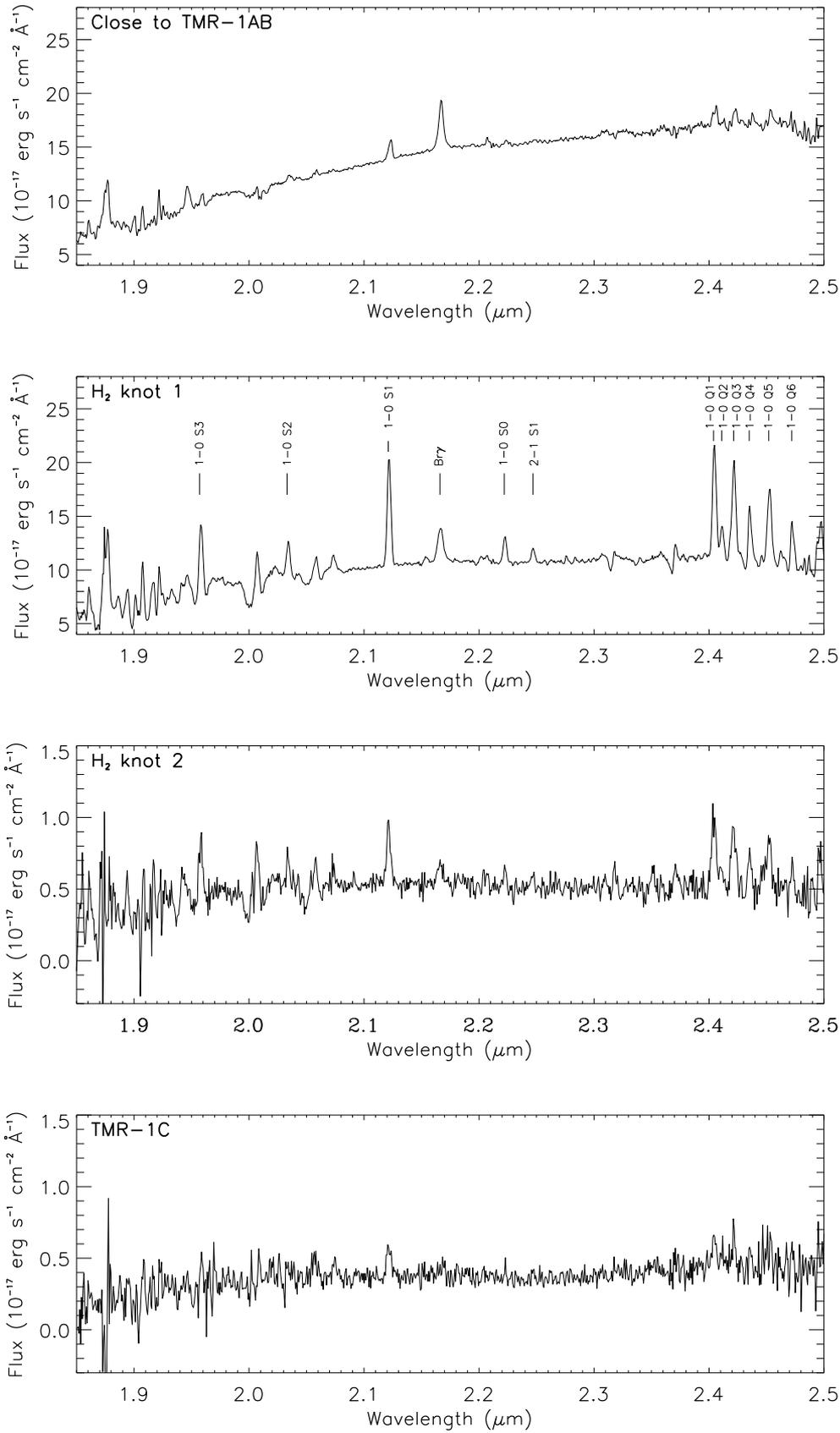}}
      \caption[]{Extracted TMR-1 spectra. All spectra cover the same wavelength range 
       from 1.85$\mu$m-2.5$\mu$m. The identification and positions of ro-vibrational H$_2$ lines 
       and the Br$\gamma$ emission line are indicated in the spectrum of the H$_2$~knot~1.
      Note that atomic hydrogen emission lines
      (in particular the Br$\gamma$-line) are
      caused or reinforced by division through the standard star, which shows
      hydrogen absorption (see text for further discussion).}
         \label{spectra}
   \end{figure*}

In order to correct for telluric absorption, as well as to attempt
flux-calibration, the spectrum of a B6IV standard star (BS3672)
was observed directly after TMR-1, with the same instrumental setting
and in the same manner (nodding along the slit) as the science
spectra. After the basic reduction the standard star spectrum was divided by a black
body spectrum corresponding to the spectral
type of the star from what the absolute spectroscopic response
was derived. 
The extracted TMR-1 spectra were then divided by the spectroscopic response
function. Absolute flux calibration based
on the spectroscopic observations of the standard star had shown to be difficult, because   
it relied on the assumption that all flux from the standard star was inside the slit. 
Moreover, the seeing worsened by a factor of $\sim 1.5$ from the time
of observation of TMR-1 to the standard star observation.
Consequently, the spectra were calibrated to match the flux density of TMR-1C as determined from 
its photometric measurement (Sec.~\ref{photom}).
Wavelength calibration was made using the OH sky emission lines and telluric 
absorption lines.
The estimated accuracy in wavelength calibration
is on the order of one pixel, i.e.\ $\sim 7$\AA.  All final four spectra are 
plotted in Figure~\ref{spectra}.

The following additional uncertainties affect the resulting spectra.
Since the standard star spectrum has not been corrected for hydrogen
absorption lines, these lines appear in emission in the TMR-1 spectra
after division by the response function.
In the case of the spectrum close to TMR-1AB  and the spectrum H$2$~knot~1, 
however, Br$\gamma$ emission is
also seen in the raw spectrum, while no emission is found in the spectrum 
H$2$~knot~2. Therefore, the Br$\gamma$ line in the first two spectra  
of Figure~3 is overestimated.
In order to roughly estimate the actual intensity of the
Br$\gamma$ (2.16$\mu$m) emission line in these spectra relative to
the 2.12$\mu$m emission line, the response
function was corrected for Br$\gamma$ absorption through a simple interpolation
over the region $\lambda$[2.15$\mu$m,2.175$\mu$m].
This way we estimate that the actual intensity ratio between
the Br$\gamma$ and the 2.12$\mu$m emission line is $\sim 0.7$
in the spectrum close to TMR-1AB and $\sim 0.14$ and the H$2$~knot~1 spectrum.
Absorption of the Earth's atmosphere does significantly affect
the quality of the spectra at and below 2$\mu$m, and also to some extent
above 2.4$\mu$m. For instance,
below 2$\mu$m the spectra are noisier by factors of 4--18
compared to the spectral
region $\lambda>2\mu$m.

Fortunately, a large number of emission lines that we identified
in the TMR-1 spectra arise from ro-vibrational transitions of
molecular hydrogen, which occupy regions
of the spectra that are not affected by telluric absorption nor by hydrogen
absorption lines of the standard star. Therefore the spectral signatures
are well measurable. The relative line fluxes of identified emission lines, normalized to the H$_2$ 1-0 S(1) flux, 
have been measured and are listed in Table~\ref{lines}.


\section{Results and discussion}

\subsection{Morphology of the nebular structure, new features and detections}
It is well known from previous observations that 
TMR-1 is associated with bright extended
nebulosity consisting of mainly two, a north-western and
a south-eastern structure. Our ISAAC images show these structures in great detail
(Figure~\ref{K_full_field}b, c).
Each of the two main nebulosities show a filament arm extension, with
the filament arm of the north-western nebulosity being more diffuse than the sharp filament
of the south-eastern structure. At the end of each filament arm we detect respectively a point-source, 
TMR-1C and TMR-1D. Another patch of diffuse emission, with an almost rectangular shape,  is seen
in the Ks-image, located $\sim 10\arcsec$ north of TMR-1AB. It is significantly less bright in the H-image,
suggesting that this structure is more highly extincted.

Considering that TMR-1 is a young stellar object during an early evolutionary state, the extended 
near-infrared emission is interpreted as scattered light from circumstellar dust agglomerations within a dusty 
proto-stellar envelope and from patches of molecular cloud material remaining from the proto-stellar collapse.
Indeed, using near-infrared polarimetric observations Whitney et al.\ (1997) report
a high degree ($\sim$50\%) of polarized K-band emission from both of the main nebulosities
and parts of the filaments,
indicating that light from the central source TMR-1AB is scattered off from them, and thereby producing
most of the emission detected in the H- and K-band images.
The observations imply that the material of the NW and SE nebulosity must be physically 
close to TMR-1AB, because otherwise they would not appear in scattered light. 
A bit surprisingly,
the NW nebula appears much fainter at 1.65$\mu$m than at 2.16$\mu$m and also slightly
redder than the SE nebula. 
We determine H-K $\approx2.6$\,mag for the NW nebula and H-K $\approx2.45$\,mag for the SE
nebula (dominated by the source TMR-1AB). A large clump of dense gas that sits partly in front of the 
NW nebulosity, detected in the CS molecular line observations of Ohashi et al.\  (1996), is the most
likely cause for the additional reddening.

However, the overall picture of TMR-1's protostellar envelope+disk+outflow structure is quite 
complex and appears undetermined, despite the availability of 
a number of observations at various wavelengths. What seems confirmed is that TMR-1AB currently 
drives a molecular outflow that
had been detected at different molecular lines, with an outflow position angle of
$\sim170\pm10^{\circ}$ (Hogerheijde et al. 1998, Terebey et al.\ 1990). Red-shifted 
gas is found north of TMR-1AB, and the NW nebulosity appears associated with lower
velocity red-shifted gas. Thus, the NW nebulosity and its filament has been interpreted as part of 
an outflow cavity. A cavity opening angle of $\sim30^{\circ}$ was derived from multi-wavelength envelope
modeling by Furlan et al.\ (2008). There is no clear detection of a blue-lobe of the outflow, which would
be expected south of TMR-1AB. 
Instead, most of the blue-shifted gas seems peaked at or close to TMR-1AB, that is roughly at the 
center of the SE nebulosity. The SE nebulosity, therefore, has been suggested to trace the 
gravitationally bound circumstellar or circumbinary material close to the proto-stars. The filament pointing
towards TMR-1C, on the other hand, could be part of an outflow cavity rim.

Adding to the complexity, we clearly observe in our deep ISAAC images a striking apparent 
symmetry between the NW and SE nebulosities,
the extending filament arms and the presence of a point-source on either end of the 
filaments (cf.\ Figure~\ref{K_full_field}b, c). The center of symmetry is determined by a 
large ``gap'' or lane of $4\farcs5$ in size, i.e.\  almost 700\,AU, in
between the two nebulosities (see also, Terebey et al.\ 1990, Itoh et al.\ 1999). While in some 
previous studies the 
dark lane had been interpreted as an extinction structure, we suggest that the gap is rather devoid of high
density material, because neither cold dust emission maps from either single dish or interferometer observations
nor other high column density tracers detect the dark lane structure  (Motte \& Andr\'e 2001, 
Petr-Gotzens et al.\ 2002, Walsh priv.\ comm.). 
The same conclusion was reached by Whitney et al.\ (1997) based on the relatively blue H-K colour observed
towards the gap structure. TMR-1AB appears offset from the gap. This may be a geometrical effect,
if TMR-1AB would be surrounded, for example, by a large ring of dense gas with an inner hole and the ring being tilted
towards the observer. The nebulosities would trace to some extent the optically thick surface 
of such a ring. But without more detailed information on the gas kinematics and observations at multi-wavelengths,
no firm conclusion can be made.

In the context of our study of the nature of the point-source TMR-1C, yet the most noteworthy observation
revealed by our images is the symmetric location of TMR-1C and TMR-1D, each at the end of a long filament 
structure. We interpret the morphology as such 
as if the filament arms had been swept up by a common symmetric outflow. The filaments are not an outflow
or jet by themselves, because they are composed of large amounts of continuum emission.
Maybe the expansion of a powerful bipolar outflow had triggered the collapse of small low-mass cores, 
leading to the formation of TMR-1C and 1D?
Based on our ISAAC images, we measure 
a position angle of $\sim 145^{\circ}$ for a line tracing roughly the direction of the filaments and
connecting TMR-1C with TMR-1D. 
The position angle of the currently observed molecular outflow, detected at several different molecular lines 
has been determined to $\sim170\pm10^{\circ}$ (Hogerheijde et al. 1998, Terebey et al.\ 1990).
One may speculate that the main outflow direction has changed counter-clock wise over
the past, because the current outflow's axis seems different from the position angle formed by
TMR-1C and 1D. Or alternatively, one binary component's outflow has ceased already and we are currently
only witnessing the other component's outflow. 
In this context we note another intriguing new detection from the deep ISAAC image:
clearly visible in Figure~\ref{K_full_field}a, detected at $1.2\arcmin$ distance south-southeast of TMR-1, 
there is a bright arc-like structure of likely swept-up material, whose morphology resembles that of
a bow-shock. However, no excessive line emission at 2.12$\mu$m is detected, which should be expected 
if the emission would be caused by shock-excitation. The {\it Spitzer}/IRAC images at 3.6mic and 4.5mic
confirm the structure (Figure~\ref{irac3_6and4_5}), but neither the IRAC images nor our deep ISAAC images
show an associated north-northwest counterpart. This is unlikely due to increased extinction, because
there is no indication for this, neither from star counts nor from dust continuum maps. Although, clumpy extinction
on small scales cannot be excluded. If a counterpart
feature exists, its emission must be faint.
The arc structure may represent a fossil, compressed  post-shock region, i.e. a signature from an
earlier passage of a shock wave. Its position angle, measured  from TMR-1AB to the arc head
is $\sim160^{\circ}$. In Sec.~\ref{formation_theo} we will follow-up on the 
discussion of possible formation mechanisms for TMR-1C and TMR-1D, and for the filament
structures.
Given these new findings we reconsider in the following sections the background star versus very low-mass
object debate for TMR-1C.

\subsection{TMR-1C, a background star ... ?}
Since TMR-1 is embedded in a region of the Taurus molecular cloud which is
heavily obscured by dust,  we test the hypothesis that TMR-1C is an
unrelated star, located
behind the molecular cloud and having its light extincted by A$_v > 20$\,mag.
Low-resolution spectroscopic observations of Terebey et al.\ (2000)
actually showed that the near-infrared spectrum
of TMR-1C is not inconsistent with an extincted background dwarf star spectrum.

Our ISAAC spectrum of TMR-1C is shown in Figure~\ref{spectra}. Its signal-to-noise ratio is 
about 8--10. The interpretation is not straightforward, as the spectrum lacks any 
spectral features
which could be useful for diagnostics. The emission line seen at 2.12$\mu$m, and
those
at wavelengths beyond 2.4$\mu$m are residuals from molecular emission from the
tip of the filament (H$_2$knot2) and are not originating from TMR-1C itself. 
The ISAAC spectrum of TMR-1C is very similar to the spectrum obtained at Keck by 
Terebey et al.\ (2000), and hence does not provide any additional new weight to the background star
origin of TMR-1C. 

%
   \begin{figure}
      \resizebox{\hsize}{!}{\includegraphics{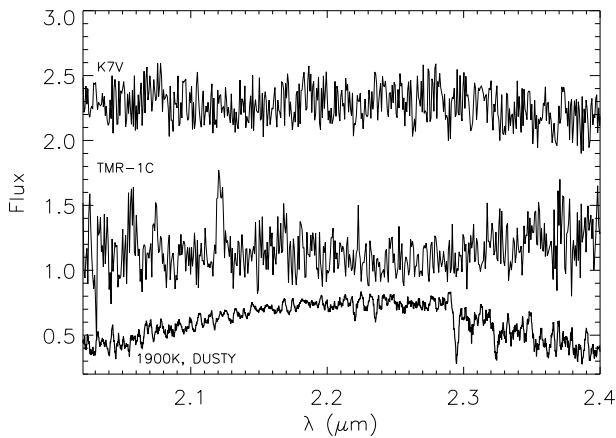}}
      \caption[]{{\it Upper spectrum:} K7\,V star spectrum from the stellar spectral
library of Pickles (1998), smoothed and extincted by A$_{\rm v}$=23\,mag, and noise added. 
{\it Middle spectrum:} TMR-1C.
{\it Lower spectrum:} A T$_{\rm eff}=1900$\,K synthetic spectrum with log~g=3.5 from the 
DUSTY model (Chabrier et al.\ 2000, Allard et al.\ 2001).
Spectra are in arbitrary flux units, with the K7\,V spectrum and the DUSTY 
model spectrum being offset for comparison.}
         \label{fittmr1c}
   \end{figure}
%
In Figure~\ref{fittmr1c}, we compare our TMR-1C spectrum with an extincted 
K7V star spectrum from the spectral library of Pickles (Pickles 1998).
The spectral resolution of the library spectra
was degraded and smoothed to fit the resolution of the ISAAC spectrum, and
noise was added to resemble the S/N of the spectrum of TMR-1C.
The amount of extinction to be applied to the dwarf star spectrum was constrained
by the $H-K$ color of TMR-1C. 
With $H-K=1.65\pm0.1$, as implied by our ISAAC photometry and from measurements 
reported in T98, A$_{\rm v}$ was constrained to 
20-27\,mag for normal interstellar dust extinction of the stellar photosphere's of
B to early-M type dwarf stars. Applying an extinction equivalent of Av=23mag to the K7V 
library star spectrum, leads to a reasonable conformity with the spectrum of TMR-1C as
shown in Figure~\ref{fittmr1c}. But note, that the comparison was mainly based on the
slope of the spectra, because no spectral features are present in TMR-1C's spectrum. As
normal stars earlier than M-type don't show a significant difference in their
spectral slope over the wavelength range analysed here, the spectral type
of TMR-1C, assuming a background dwarf, can only be classified as M5 or hotter,
and the K7V star spectrum that we show in Figure~\ref{fittmr1c} just gives an example of a conceivable fit. 

   \begin{figure}
      \resizebox{\hsize}{!}{\includegraphics{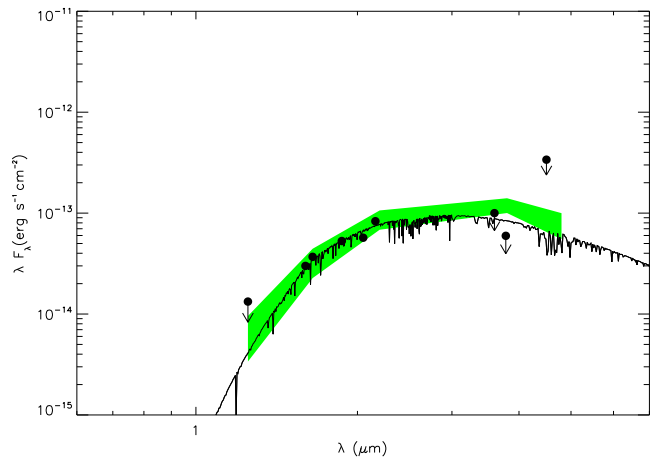}}
      \caption[]{Observed spectral energy distribution of TMR-1C (filled black circles, based on Table~\ref{phot}
and on values reported in T98; filled black circles with down-heading arrows indicate upper 3$\sigma$ flux limits).
The black solid line represents the spectral energy distribution of a NEXTGen stellar model source 
(Hauschildt et al. 1999) with T$_{\rm eff}$=4000K, log~g=5.0 and extincted by A$_V = 23.0$mag. 
The shaded band indicates the range in flux distribution expected for objects of 3-7M$_{\rm Jup}$
at the age of 1\,Myr and seen behind A$_{\rm V}$=10-22\,mag of extinction, that are overall consistent with 
the observed SED of TMR-1C for $\lambda < 2.5\mu$m. }
         \label{sed}
   \end{figure}

More information on the nature of TMR-1C can be obtained from an analysis
of its full spectral energy distribution, as much as it is accessible. To do so, we
use our photometric measurements and photometry from T98 in order to
compile the spectral energy distribution of TMR-1C.
Then, we compare to model spectra using NextGen stellar atmosphere
models (Hauschildt et al.\ 1999), and find that {\it for  wavelengths shorter than 2.5$\mu$m},
the flux distribution of TMR-1C is, for example, well represented by a main-sequence dwarf star with
T$_{\rm eff} = 4000$\,K, seen through A$_v=23$mag of interstellar extinction  (Figure~\ref{sed}).
This is consistent with the result shown in Figure~\ref{fittmr1c}. As mentioned already 
above, combinations of lower values of A$_v$ and T$_{\rm eff}$, or higher values
of A$_v$ with higher T$_{\rm eff}$ do also lead to a good fit. With an upper limit of  A$_v=30$mag
a stellar photosphere with T$_{\rm eff}$ as high as 20000\,K is still in agreement with the
observed spectral energy distribution in the near-infrared regime.
However, the obviously strong decline of TMR-1C's
flux distribution in the mid-infrared, as witnessed
by its non-detection, is inconsistent with any reddened normal stellar photospheres. 
We conclude that TMR-1C is not an extincted background star.

Our conclusion  is also supported from a purely statistical
point of view, because the chances
of TMR-1C being a background object projected along the line 
of sight close to TMR-1AB are low: we calculate the probability that a background object
coincidently falls within $10\arcsec$ to TMR-1AB, which is the separation between TMR-1C and 1AB.
For doing so, we first determine the stellar density of K-band sources from our 
relatively large field ISAAC observations, assuming that all faint objects are background sources. 
Since we detect 
25 stars with a Ks-magnitude similar or fainter than that of TMR1C,  we derive
$\sim 1\times 10^{-3}$stars/arcsec$^2$, and hence a probability of 0.22 to find a background object as 
close as $10\arcsec$ to TMR-1AB.  But TMR-1C is not only found somewhere at $10\arcsec$
distance around TMR-1AB, but at the tip of a narrow filament. We note that the 2D spectrum clearly 
reveals that the molecular hydrogen emission from the filament exactly
stops where TMR-1C is located. In other words, the statistical evaluation of the hypothesis 
that TMR-1C is an unrelated background object, needs also to add
the probability for finding the background object at a specific direction (i.e. in the direction of the
filament) around TMR-1AB, which we divide in segments of 1 degree.
Then, this probability to find a random background 
star as close as $10\arcsec$ to TMR-1AB {\it and} within $1^{\degr}$
of a certain position angle drops to $6\times 10^{-4}$.

\subsection{... or a very low-mass object?}
\label{vlm}
If TMR-1C is not a chance projected background star, what kind of object is it?
Since the spectrum is obviously pure continuum it is not
consistent with a bullet nor dense knot of shocked molecular gas. 
Moreover, at all imaging observations reported so far, TMR-1C appears 
as an unresolved point-source; the HST image even shows the first Airy diffraction ring. 

The study of T98 proposed that 
if TMR-1C is physically associated with TMR-1AB it should be a young sub-stellar object,
since it is several magnitudes less luminous than TMR-1AB, which is a proto-stellar
system of $\sim 0.5M_{\sun}$ (Kenyon et al.\ 1993, T98). TMR-1C's bolometric
luminosity could be as low as $10^{-3}-10^{-4}{\rm L}_{\odot}$ as estimated by T98 and 
Terebey et al.\ (2000).
In Figure~\ref{sed} we plot, as a shaded band, the range of SEDs for planetary mass objects
with 3-7 M$_{\rm Jup}$ and of the age 1\,Myr (based on DUSTY and COND sub-stellar evolutionary
models of Chabrier et al.\  2000, Baraffe et al.\ 2003) that are consistent with the 
near-infrared spectral energy distribution  
of TMR-1C. Depending on the mass of the object, and hence on the
effective temperature, different values of extinction, ranging from Av=10-22mag had to be added to achieve a fit with
the near-infrared part of the observed SED of TMR-1C.
While the general slope of TMR-1C's near-infrared spectral energy distribution is 
consistent with extincted photospheres of 3-7 M$_{\rm Jup}$ mass objects, the
expected value of effective temperatures for such low-mass objects ($< 2100$K),  however, is in strong contradiction 
with our observed K-band spectrum that indicates higher T$_{\rm eff}$ ($ \gtrsim 3000$K).
K-band spectra of cool objects with effective temperatures  $<$ 2500K 
display prominent signatures in the K-band, like
deep methane and/or water vapor absorption bands, the $v=2-0$ vibration-rotation
bands of CO at $\lambda > 2.29\mu$m, and the NaI atomic absorption line 
at 2.21$\mu$m (Burrows et al.\ 2001, and references therein). 
In general, the depth of the H$_2$O and CH$_4$ absorption increases from 
M-class to L- and T-class objects, though at some phase during L 
spectral class H$_2$O absorption is weakened by dust formation and 
re-radition. But in any case, the molecular absorption is such pronounced, 
especially for the coldest very low-mass T-dwarfs, that it is easily detectable
even in low signal-to-noise spectra (e.g.\ Cuby et al.\ 1999).
It further seems that these strong absorption bands
are also present in young ($\leq 1$Myr), low gravity, low-mass
objects. For instance, distinct water vapor absorption bands have been 
observed in near-infrared spectra of brown dwarfs and planetary mass 
objects in the very young $\rho$ Oph and Orion Trapezium cluster
(Wilking, Greene \& Meyer 1999; Lucas et al. 2001; Lucas et al.\ 2006).
None of the features described above are present in our spectrum of TMR-1C.
In Figure~\ref{fittmr1c} we show for comparison the synthetic spectrum of 
a young 5\,M$_{\rm Jup}$ object having T$_{\rm eff}=1900$\,K and log~g=3.5.
This model spectrum, generated by the Lyon group based on the Ames-DUSTY stellar evolutionary
model (Chabrier et al.\ 2000, Allard et al.\ 2001), is clearly different from the spectrum of TMR-1C.
Furthermore, as implied by Figure~\ref{sed}, any young planetary mass object
should be brighter than our measured upper detection limits at mid-infrared wavelengths.

   \begin{figure}
      \resizebox{\hsize}{!}{\includegraphics{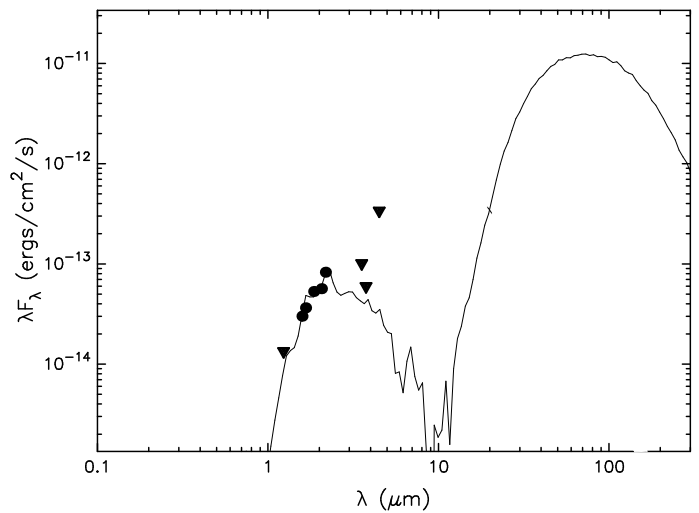}}
      \caption[]{Spectral energy distribution of TMR-1C (indicated by  filled dots,
      triangles represent upper limits) compared to a model SED for a central source of 0.13${\rm M}_\odot$ with a
      circumstellar disk inclined to our line of sight by $i=87\deg$, an outer disk radius of 94\,AU, an age of
      $\approx 4\times10^5$yr, and located
      at a distance of 138\,pc seen through an interstellar extinction of A$_V$=18.0mag. }
         \label{robi_model}
   \end{figure}  

An alternative solution may be that TMR-1C is physically associated with TMR-1AB, but instead of
being an extremely low-mass object, it may be a slightly higher mass object having its light significantly 
dimmed by a surrounding circumstellar disk seen edge-on.  
In this case, the emergent flux in the near-IR is dominated
by scattered light from the disk surface.
TMR-1C, if occulted by a disk, would appear subluminous for its spectral
type and could exhibit an observed near-IR spectral energy distribution  
different from the object's intrinsic colours. This hypothesis for the nature of TMR-1C was actually 
already mentioned in an earlier paper by Hartmann et al.\ (1999).
Direct and indirect evidence for the presence of disks
around young substellar objects have been found by several authors
(e.g.\ Natta et al.\  2002, Luhman et al.\ 2007). Such studies show that
disks around young brown dwarfs have properties which are, in general, 
similar to those derived for circumstellar disks of low-mass stars.
Also, low-resolution near-IR spectra of objects with edge-on
disks often appear featureless (e.g.\ Brandner et al.\ 2000, Scholz et al.\ 2008), as it is
the case for the TMR-1C spectrum.
To explore this possibility, we used the SED model fitting 
tool\footnote{http://caravan.astro.wisc.edu/protostars/sedfitter.php} by Robitaille et al.\ (2007)
that has been developed to analyze the SEDs of young stellar objects. A particular advantage
of this tool is that it allows the inclusion of upper flux limits. In order to limit 
the solutions to models that are physically relevant for the case of TMR-1C, 
we postulate that the source's distance must be within 117-157pc, the range in distance considered for the
Taurus molecular cloud  (Torres et al.\ 2007), and that its age must be $\lesssim$1\,Myr. 
The model that then agrees best with the observed SED
of TMR-1C is shown in Figure~\ref{robi_model}. The model SED shows a steep decline beyond
2.2$\mu$m, and a strong excess at $\lambda >$10mic. The underlying model is that of a 
central source of $\sim 0.13{\rm M}_{\odot}$ at a distance of 138\,pc, and with the following 
parameters: age of the source $\sim 4\times10^5$yr, A$_{\rm v}=18$\,mag, T$_{\rm eff}$=3000K,
inclination angle of the circumstellar disk, $i=87^{\circ}$, inner disk radius, $r_i=1.8$\,AU, and disk
outer radius, $r_o=94$\,AU. 

The predicted size of the disk seems large, given that TMR-1C appears unresolved
in all the HST and ISAAC images. An edge-on disk as large as $r_o=94$\,AU and its 
associated bipolar scattered light structures should have appeared resolved in the high spatial
resolution HST images (FWHM$\sim 0\farcs15 - 0\farcs2$) and possibly also in the best ISAAC images.
The fact that TMR-1C appears as a point-source implies that the actual disk must be smaller than $\sim$20 AU.
A decreasing outer disk size will
mostly affect TMR-1C's spectral energy distribution longward of 10$\mu$m, i.e.\  in a regime that is 
unconstrained due to the lack of observations. In a picture where TMR-1C had been ejected in a
dynamical interaction within the
small proto-stellar cluster of TMR-1, it would be expected that any primordial disk of TMR-1C has
been truncated (cf.\ Sec.~\ref{formation_theo}). 
We note that if one allows models for sources
at slightly larger distances (up to $\sim$170pc) or slightly older objects (up to $\sim1.2\times10^6$yr), 
one finds that 
also objects with $\sim0.2-0.4{\rm M}_{\odot}$ having disk sizes as small as 19\,AU provide reasonable fits
to the SED of TMR-1C.
But since the SED is unconstrained beyond 5$\mu$m, it is currently not possible to distinguish
any further between the model solutions. SED models for objects with 
central source masses $<0.1{\rm M}_{\odot}$ are not available. However, an important point 
we wish to emphasize,
is that all models that provide a good fit imply a circumstellar disk with high 
inclination angle of $i>80^{\circ}$.

Observational evidence for an edge-on disk surrounding TMR-1C would be an infrared 
excess emission at $\lambda>10\mu$m. Unfortunately, the spatial resolution of available observations
at mid to far-infrared wavelengths are too low to disentangle a potential emission
from TMR-1C from the emission of TMR-1AB.  Other evidence, as discussed above,
could be provided by near-infrared images with high sensitivity
and a spatial resolution better than $0\farcs1$ that should reveal any resolved
bipolar scattered light structure.

\subsection{On the nature of the filament}
It has been suggested that the arc-shaped filamentary structure 
that seems to 'connect' TMR-1C with TMR-1AB is a 
material tail formed during an encounter of proto-stellar 
disks surrounding TMR-1A and B respectively (Lin et al.\ 1998, T98). 
The same encounter may have caused
the formation of TMR-1C from fragmentation of a part of the filament into
a very low-mass object, or through ejection during the dynamical interaction.
As such there would be a clear physical relationship between the 
filament and TMR-1C.

\begin{figure}
      \resizebox{\hsize}{!}{\includegraphics{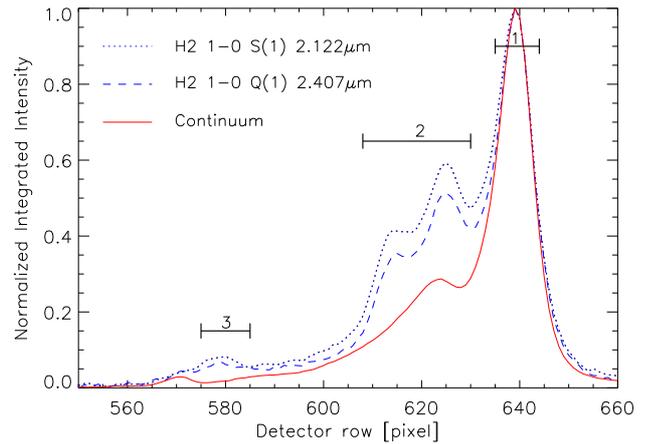}}
      \caption[]{Spatial distribution of the intensities in the H$_2$ emission
                lines (dotted and dashed blue lines) along with the continuum
                (red solid line). The intensities have been normalized to their 
                peak emission in the ISAAC TMR-1 spectrum. The apertures of extracted spectra 
                are also indicated: (1) close to TMR-1AB, (2) H$_2$~knot~1, (3) 
                H$_2$~knot~2. }
         \label{h2exc}
   \end{figure}

The higher spectral resolution of our ISAAC data as compared to 
the previous spectrum of the filament (Terebey et al.\ 2000) allows us 
to extract new physical information for the nature of the filament.
Previous studies mainly report on scattered light emission arising from the
filament. However, in spite of a clear component of continuum emission,
we believe that there is another 
distinct process present, that is shocks. 
In Figure~\ref{h2exc} we plot the spatial
distribution of the intensities in the H$_2$ 1-0\,S(1) and H$_2$ 1-0\,Q(1)
emission lines, and in the continuum, along the filament.
Obviously there are 2 locations where the emission in molecular hydrogen 
is significantly in excess.
Spectra at these 2 locations along the filament have been extracted 
(H$_2$~knot~1 and H$_2$~knot~2 in Figure~\ref{spectra}, note: the H$_2$~knot~1
location, i.e.\ aperture 2, actually includes 2 peaks of H$_2$ emission which showed indistinguishable 
spectra and where therefore combined to improve the signal-to-noise).
Both spectra show a wealth of emission lines, which are predominantly 
ro-vibrational lines of molecular hydrogen. 

Molecular hydrogen
emission mainly originates from either of the two physical processes: shock
excitation or ultraviolet fluorescence. We use the intensity ratio of the 
transitions H$_2$ v=1-0\,S(1) at 2.12$\mu$m
and H$_2$ v=2-1\,S(1) at 2.24$\mu$m to distinguish between these two cases. 
Thermal excitation via shocks
should be the responsible mechanism if
this ratio is $\gtrsim 5$, and gas densities are not too high, i.e.\
$\lesssim 10^5 {\rm cm}^{-3}$ (Shull and Beckwith 1982). Since we measure
a 2.12$\mu$m/2.24$\mu$m ratio of $\sim 8.25$ and of $\sim 5.6$, for the 
H$_2$~knot~1 spectrum and the H$_2$~knot~2 spectrum respectively,
and we do not observe any lines from
transitions of high vibrational levels, we conclude that the H$_2$
emission in the filament is due to shock excitation. 
Support for the shock interpretation is also gained from H$_2$ ortho-para ratios
discussed below.

Slightly different is the 
situation for the spectrum extracted close to TMR-1AB, which does not show 
any detectable emission at 2.24$\mu$m, suggesting the absence of molecular
gas at higher temperatures ($\gtrsim$ 2000\,K).
The Br$\gamma$ recombination line of atomic hydrogen at 2.166$\mu$m, on the other hand, 
is one of the most prominent lines in this spectrum. This indicates that a process
must be present that is capable of ionizing hydrogen. Winds from the young
stellar object is one of the likely responsible mechanisms. But also disk accretion onto the central
source may be the cause for the
Br$\gamma$  emission. Prato et al.\  (2009) present a K-band spectrum 
of TMR-1AB which shows similar features as our ISAAC spectrum obtained close to TMR-1AB
(remember that the ISAAC slit positioning is offset by almost $1^{\prime\prime}$ from the TMR-1AB peak emission). These 
authors determine, from the Br$\gamma$ emission line luminosity, a mass accretion rate of 
$1.9\times10^{-7} {\rm M}_\odot/{\rm yr}$ for TMR-1, which is quite consistent with Br$\gamma$  emission
being generated by accretion.

What causes the shocks in the filament? It has been known for long, that TMR-1 is associated
with a large molecular outflow (Terebey et al.\ 1990, Hogerheijde et al.\ 1998)
observed e.g.\ at $^{12}$CO (3--2). The direction of this outflow 
is in very good agreement with the direction of a jet indicated in FeII 
narrow-band near-infrared images (Petr-Gotzens et al.\ 2002). The position
angle of the filament is different by $\sim 20^{\circ}$ from the 
position angle of the outflow and jet. A plausible scenario is that  
the filament is part of the edge of a cavity cleared by a lower velocity
outflow, as for example observed in HH\,211 (Gueth \& Guilloteau 1999).
The molecular hydrogen emission knots along the filament would
then arise, because this lower velocity flow
hits into the cavity rim, thereby creating a C-type shock (see Sec.~3.4.1).
Or, the filament is intrinsic, pre-existing
dense material being shocked by the outflow as it happens to be in its way.
It is unlikely that the filament is a jet by itself. While bent
jets have been observed (Davis et al.\ 1994, Bally \& Reipurth 2001),
the significant amounts of continuum
emission coming from the filament, together with a high degree of
polarization (Whitney et al.\ 1997), and detection of high column
densities associated with the filament (Hogerheijde et al.\ 1998, Motte \& Andr\'e 2001),
exclude a jet.

\subsubsection{A shock-model for the emission from the filament}
While we could establish, that the H$_2$ emission along the filament 
is caused by shocks, the evaluation of the precise details of which shock excitation mechanism
is at work, and what is the underlying shock physics is more complex. A powerful method to analyze 
different possible mechanisms is to compare  measured Column Density Ratios (CDRs) to 
various shock excitation models in a schematic CDR diagram.
This approach was taken in this study, and we converted our 
flux measurements of the ro-vibrational molecular hydrogen lines (Table~\ref{lines})
into Column Density Ratios following the calculations as given in Eisl\"offel et al.\  (2000) 
and Smith (2000). Thus, we determine the column of H$_2$ in the upper energy level
necessary to generate each line flux. In the CDR diagrams,  we compare column density 
ratios to those of a slab of gas of constant temperature (2000K). By so removing the 
exponential Boltzmann factor, we are able to analyse the results in detail.

\begin{figure}
      \resizebox{\hsize}{!}{\includegraphics{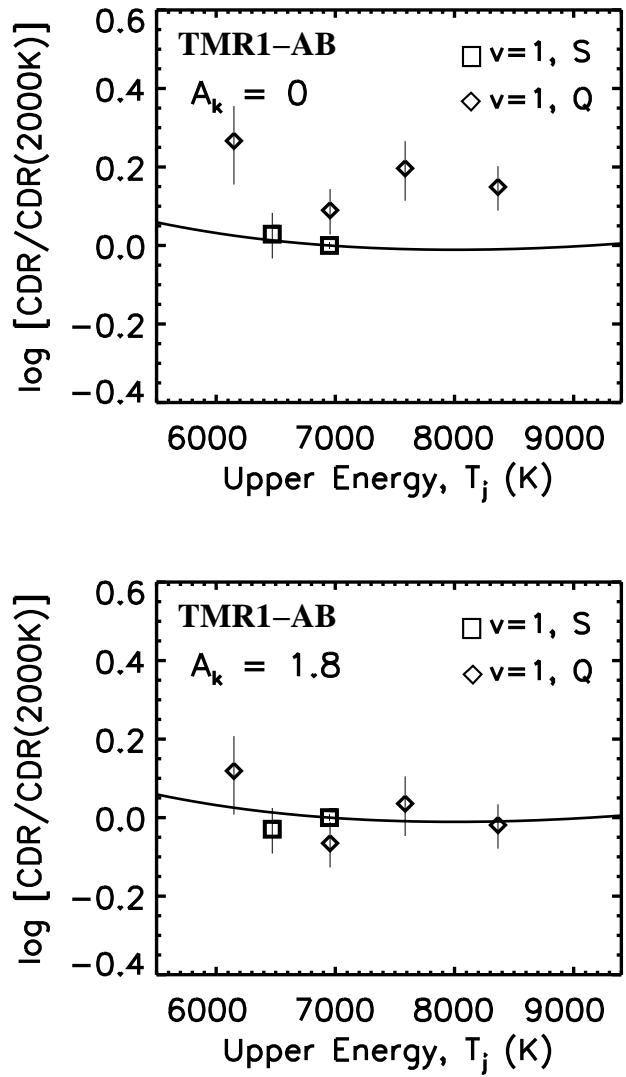}}
      \caption[]{Log(CDR) diagram derived for the position close to TMR-1AB. The solid line represents 
        a C-type bow shock model. }
         \label{cdr_spec1}
   \end{figure}

\begin{figure}
      \resizebox{\hsize}{!}{\includegraphics{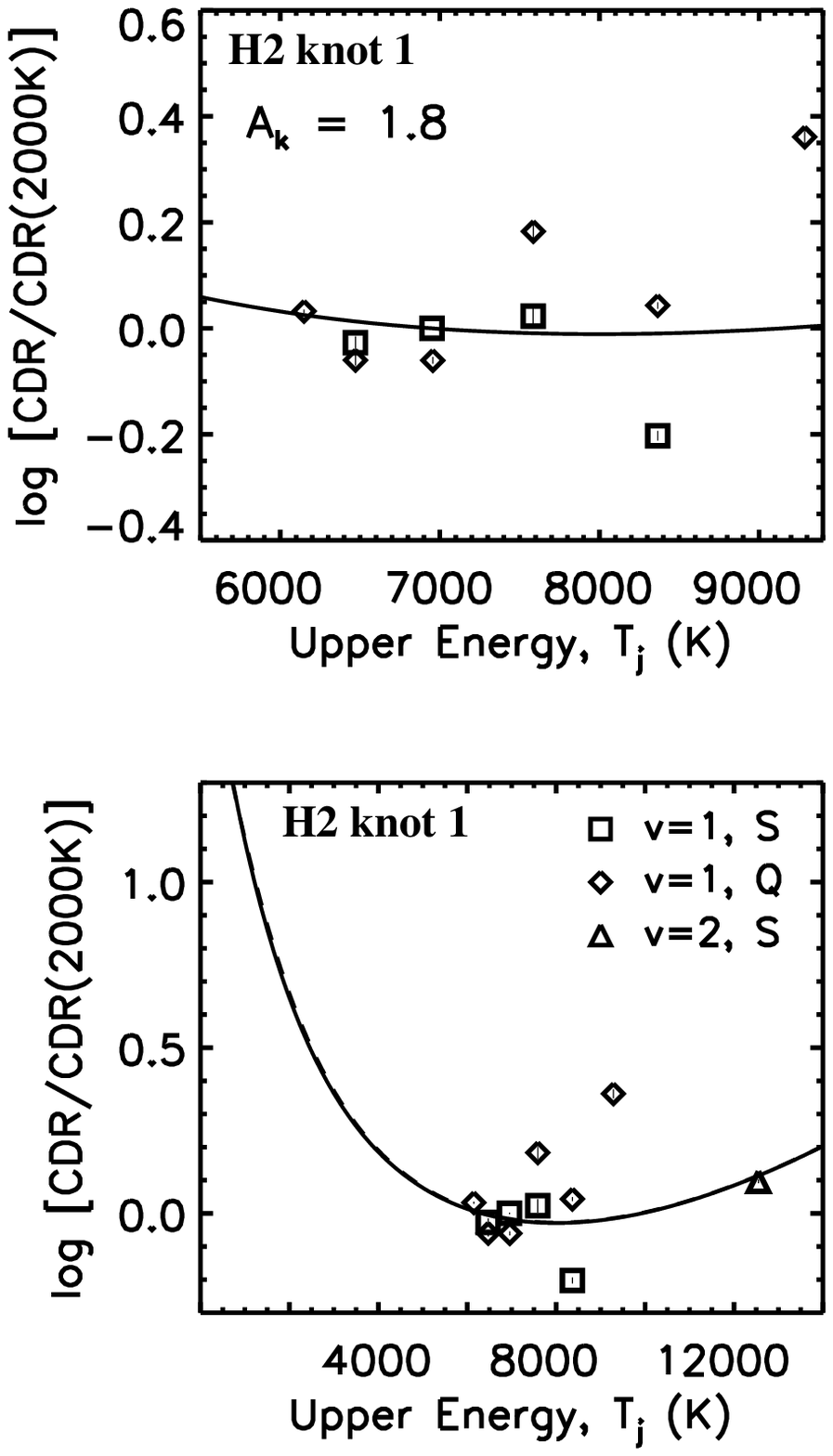}}
      \caption[]{Log(CDR) diagram derived for the H$_2$~knot~1 position.
          The solid line represents a C-type bow shock model. The upper panel and lower panel 
        show the same datapoints, with the exception that the lower panel has a larger range
        of energies, and includes the measurement at H$_2$ 2-1 S(1).}
         \label{cdr_spec2}
   \end{figure}

\begin{figure}
      \resizebox{\hsize}{!}{\includegraphics{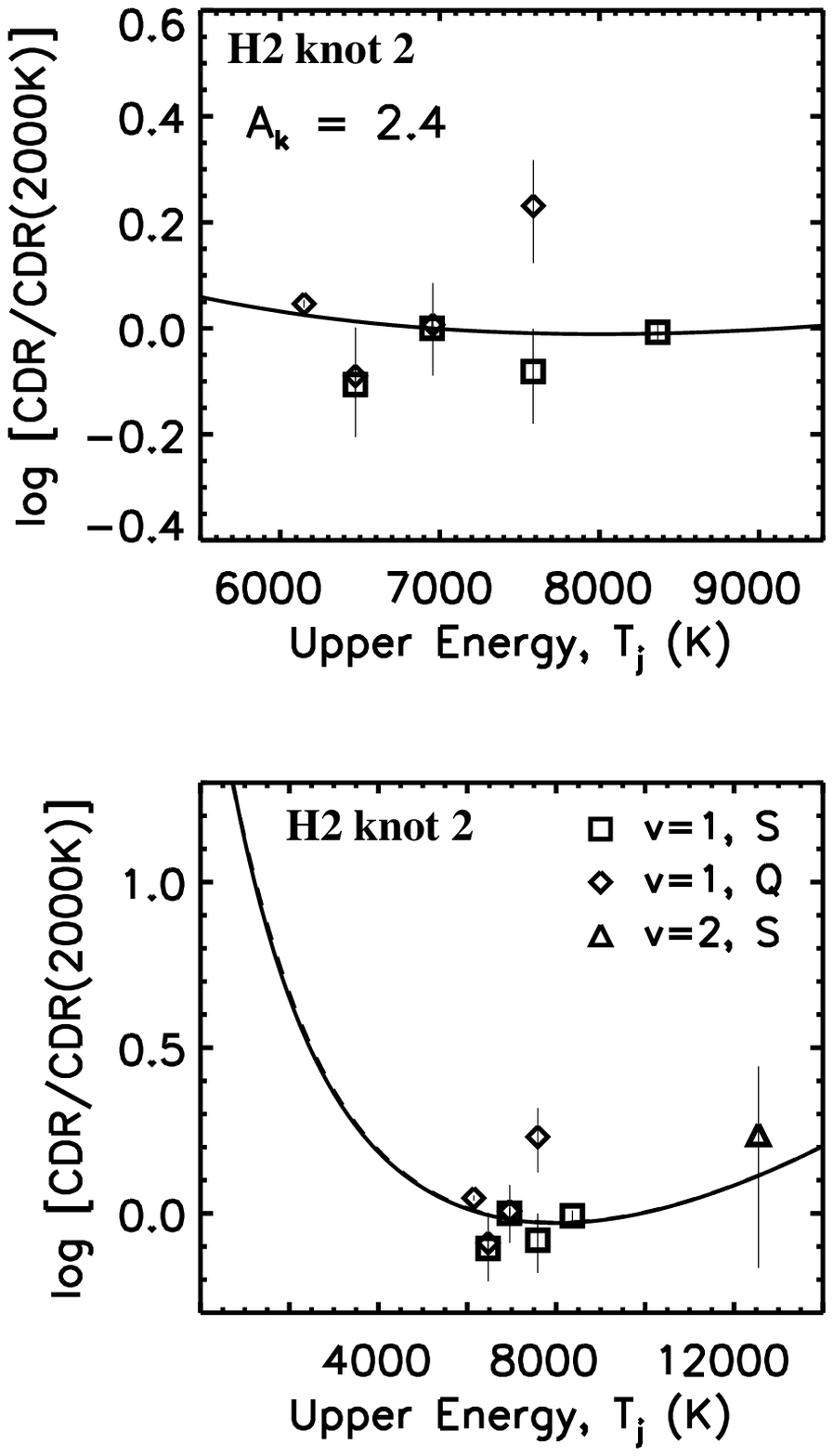}}
      \caption[]{Log(CDR) diagram derived for the H$_2$~knot~2 position, close to TMR-1C.
        The solid line represents a C-type bow shock model. The upper panel and lower panel 
        show the same datapoints, with the exception that the lower panel has a larger range
        of energies, and includes the measurement at H$_2$ 2-1 S(1).}
         \label{cdr_spec3}
   \end{figure}

The CDR diagrams for each position along the filament are shown in 
Figures~\ref{cdr_spec1},\ref{cdr_spec2},\ref{cdr_spec3}. The shock model plotted on all
diagrams is a C-type bow shock. The sparesness of the data, in particular with 
respect to the range in temperatures, and their scatter prevent detailed fits.
Note that a constant temperature model (corresponding to a linear fit) would be consistent with all the data. 
A horizontal line on the CDR diagrams would correspond to 2000\,K. A possibility is a single C-type shock in which
the emission derives from an extended zone with a quite narrow
range of gas temperatures (Smith \& Brand 1990). 

However, shock waves of all types do not maintain gas at a fixed temperature and we expect a range of
 temperatures to be present. As an example,
we demonstrate on the diagrams that at all locations a gently curved shock with some range
of shock velocities is also consistent. This issue can be resolved by observing pure rotational lines from low 
levels in the mid-infrared since very high CDR values are expected from bow shocks (see lower panels 
of the Figures~\ref{cdr_spec2},\ref{cdr_spec3}). 

All CDR diagrams display no evidence for ortho-para ratios deviating from that usually associated with shock waves. 
Fluorescent emission from unshocked cloud gas in star formation regions
yields enhanced columns of gas in the para upper energy levels (e.g. Li et al. 2008). Here, however, we find that, 
within the errors, there is no deviation from a thermalised ortho-para ratio of 3
(Smith et al. 1997), corresponding to linear fits on the diagrams (note the large errors associated with the 
noisy parts of the high Q-branch and high S-branch transitions).  

\subsubsection{Extinction along the filament}
Estimates of the interstellar extinction can be directly
derived from line flux ratios of the Q-branch at $2.4-2.5\mu$m, independent
of the specific shock model. 

In Figure~\ref{cdr_spec1} we show the data for the filament position
close to TMR-1AB with 2 different extinctions (upper and lower panel
of the figure). With A$_{\rm K}=0$mag all the Q-branch datapoints lie above the S-branch 
datapoints, while a much better fit of all datapoints with the model is achieved for
A$_{\rm K}=1.8$mag (lower panel). The uncertainty in the derived extinction is $\pm0.3$\,mag.
This value can be compared with the extinction determinations toward TMR-1AB reported in
Prato et al.\  (2009) and based on low-resolution K- and L-band spectra. They find A$_{\rm V}=(24.2\pm1.1)$mag
using an optical depth measurement of the ice absorption feature at 3.1$\mu$m, and A$_{\rm V}=(20.8\pm1.8)$mag
based on the source's  near-infrared colors.  Both extinction values are higher than our extinction derived from the 
molecular hydrogen line ratios. This is not surprising as the different methods trace different lines
of sight through the material causing the extinction.
In particular, the molecular hydrogen emission lines are supposedly produced in shocked regions away 
from the protostellar photosphere. 
Also, our ISAAC spectrum is slightly off ($\sim1\arcsec$) from the TMR-1AB protostar system, while Prato et al.\ (2009) 
centered their spectrum on TMR-1AB. 

For the filament positions H$_2$~knot~1 and H$_2$~knot~2 (Figure~\ref{cdr_spec2} and \ref{cdr_spec3}) 
the modeling is somewhat facilitated by the presence of the H$_2$ 2-1 S(1) line, enabling a 
fit over a larger range in temperature. The calculated extinction at the position of the H$_2$~knot~1
is A$_{\rm K}=1.8$mag, and hence not different from the extinction we determined closer to TMR-1AB. 
At the position of the H$_2$~knot~2,
i.e.  only $\sim1\farcs3$ away from TMR-1C, we measure an A$_{\rm K}$ of $2.4$mag 
[A$_{\rm V}=(21.4\pm2.7)$mag], which is in good agreement with the 
values of extinction estimated towards TMR-1C (cf.\ Sec.\ 3.2 \& 3.3).  
Moreover, the extinction determined at the H$_2$~knot~2 is
significantly higher than at the H$_2$~knot~1 and close to TMR-1AB, 
indicating that the extinction {\it increases}
from close to the main source along the filament. This finding seems to be in contradiction to the conclusion
reached by T98 where they observe an increasing flux ratio with decreasing wavelength
for TMR-1C relative to the protostars TMR-1AB and they conclude that there is decreasing extinction towards TMR-1C. 
However, TMR-1C could be located in front of the filament while the filament
is indeed receding into the denser parts of the molecular cloud.

If we consider the continuum in the spectrum taken close to TMR-1AB
being dominated by the continuum from the protostars, then the spectra continuum slopes  (cf.\ Figure~\ref{spectra}),
in principle, imply a lower interstellar extinction towards TMR-1C than 
towards TMR-1AB. Our SED analysis in Sec.~\ref{vlm} suggests A$_{\rm V}=18.0$mag for TMR-1C, and Prato et al.\ (2009)
derive A$_{\rm V}=(24.2\pm1.1)$mag for TMR-1AB. In addition,  
strong excess emission from accretion and the (unknown) circumstellar disk 
geometry are also contributing to the continuum spectral shape of TMR-1AB. Disentangling the different
effects is beyond the scope of this paper.

\section{Formation scenarios for TMR-1C (and TMR-1D)}
\label{formation_theo}
We now discuss different models 
that may qualitatively explain the formation of TMR-1C as a low-mass object physically associated with the
proto-binary TMR-1AB. In such a picture of the TMR-1 system, TMR-1C {\it and} TMR-1D are both very low
mass objects (either low-mass stars or brown dwarfs) and their symmetric locations on opposite
sides of the main source TMR-1AB shall be taken into account when considering possible formation
scenarios.

A scenario worth debating is that of outflow triggered star formation (Sandell \& Knee 2001).
This mechanism naturally yields collapsed objects within the high-pressure region associated with a jet or outflow impact.
Suppose a string of well-spaced pre-stellar cores existed in this region before one of them collapsed to form the binary system. 
Provided that the developing proto-stellar jets are directed along the axis connecting the cores, the subsequent impact could 
trigger a collapse in adjacent cores simultaneously on either side of the expanding outflow. Once the low-mass core
is forced into collapse the formation of a low-mass object proceeds 'as normal', i.e.\  its mass continues to grow 
via disk accretion. 
This model has the advantage of generating symmetric collapsed objects on opposite sides at the locations of adjacent cores in the string.
Moreover, the on-going outflow may even drive away extended disk and envelope material, generating swept-up dust ridges and shells.
The filament would represent denser material along the outflow direction, possibly consisting of swept-up matter,
that scatters light having escaped from the main source TMR-1AB through a low density cavity excavated by the jet/outflow. 
At the densest knots along the filament the outflow impact gives rise to shock excited emission. Low-mass pre-stellar
cores may also be produced in the shock-compressed layers of swept-up dusty filaments (Padoan \& Nordlund 2004).
Sandell \& Knee (2001), for example, detect several compact mm-sources in a curving arc of dust near the proto-stellar
object SSV13 in the NGC1333 star-forming region. 

Intriguingly, the protostellar outflow HH\,211 also possesses two symmetric star-like objects at the tips of the presently visible outflow 
(see Fig.\,1 of Eisl\"offel et al. 2003). As with the TMR-1 outflow, the near-infrared H$_2$ outflow of HH211 raises curiosity because
it also contains a filament of material with strong continuum emission (O'Connell et al.\ 2005).  

Another possibility to discuss is the formation of TMR-1C (and TMR-1D) from disk fragmentation processes.
Numerical simulations recently published by Stamatellos \& Whitworth (2009) show that a number of
very low-mass objects with masses in the range 5-200\,M$_{\rm J}$ can be produced by gravitational disk fragmentation
occuring in an extended circumstellar  disk.
Most of the produced objects are brown dwarfs ($>$13\,M$_{\rm J}$) and low-mass hydrogen-burning stars. 
Already $\sim10^4$yrs after the formation of these objects from disk fragmentation, dynamical N-body interactions 
lead to an ejection of most them, i.e.\ making them unbound and liberate them into the field, but a few objects typically
remain bound. Quite interestingly,
the most common (i.e.\ the most likely) configuration resulting from the simulations after a few hundred thousand
years is that of a solar-like primary with a nearby low-mass stellar companion ('TMR-1AB'), and two wide (r$>$100AU) companions,  
consisting of a low-mass star ('TMR-1C') and a brown dwarf ('TMR-1D'); a
configuration that very closely resembles what we believe we observe for the TMR-1 system. In this model, TMR-1C and TMR-1D 
remain weakly bound to TMR-1AB. They are also expected to retain their circumstellar disks,
although at relatively small sizes, as the disks may had been stripped off even further during dynamical interactions.
Dynamical interactions are also held responsible for inhomogeneous orientations of the young objects' circumstellar disks. 
Non-coplanarity is frequently observed in the modelling (Stamatellos \& Whitworth 2009).
It seems plausible that a close passage of, e.g.\ TMR-1C to TMR-1AB, early in the evolutionary history of the system, 
caused the circumstellar disk and hence the jet of TMR-1A (or B) to precess, which is what our observations of a 
bow structure detected at $> 1^{\prime}$ distance from TMR-1 indicate.

A shortcoming of the 'disk fragmentation with follow up ejection' scenario for the origin of TMR-1C and D is that it does not explain
the existence of symmetric filaments and 
a physical relation of those to TMR-1C and TMR-1D. Although filamentary patterns of the
fragmenting primordial disk are observed in the simulations, they are usually short-lived and much smaller in
size than what is observed for TMR-1. Therefore, we consider for a moment that 
both, primary and secondary component of TMR-1AB originally
condensed out of the collapsing pre-stellar core, and that both were accompanied by
their own extended protostellar disk. Presume that in a subsequent encounter of the systems, tidal structures 
consisting of disk material developed ('the filaments'), and that one proto-stellar component was captured by the 
more massive component to form the 
proto-binary TMR-1AB. Naturally, the tidal filaments form symmetric structures. 
As frequently observed in different numerical simulations, very low-mass objects are then formed out of the densest
parts of the filamentary tidal arms (Shen \& Wadsley 2006, Lin et al.\ 1998), often far from the main source. This way
there would be a clear causal dependence for the filaments and the candidate low-mass objects TMR-1C and 1D.
Shen \& Wadsley (2006) further show that the formed 'proto brown-dwarfs' posses circumstellar disks of sizes up to $\sim$20AU.
The objects have typical masses in the range $2-73{\rm M}_{\rm J}$, but the simulations are stopped before mass grow
through disk accretion could be completed, so that the objects' final masses may be slightly
higher.
 
\section{Summary and Conclusions}
In this paper we have used VLT/ISAAC infrared imaging and spectroscopy data, complemented with
near-infrared HST and {\it Spitzer/IRAC} photometry in order
to study the nature of low-luminosity objects detected in the close surrounding of the 
young stellar object TMR-1 (actually a binary source: TMR-1AB). We have also discussed the 
particular morphology 
of the dust distribution around TMR-1AB as it appears in the near-infrared images, and how the
specific dust structures may be related to a low-mass star formation process.
Our main objective has been to add further constraints on the potential sub-stellar nature of the
faint source TMR-1C, which is a wide companion to the young proto-stellar binary TMR-1AB.
The striking location of TMR-1C at the tip of a narrow filament emanating from the nebulosity surrounding
TMR-1AB suggested a physical association and implied a substellar nature for TMR-1C in previous studies (T98).
Our most important results derived from the analysis in this paper are summarized as follows:
\begin{itemize}
\item The  near-infrared K-band spectrum of TMR-1C is
featureless, and fully consistent with previous lower resolution spectra. The spectrum indicates
that TMR-1C's effective temperature is $\gtrsim 3000$K and rejects the
possibility of a very cold low-mass object. 

\item The near to mid-infrared spectral energy distribution of TMR-1C clearly
shows a decline in flux for $\lambda > 2.5\mu$m. It can not be fitted 
with an extincted background dwarf star,  nor
with a brown-dwarf or planetary mass object photosphere,
for which the object should have been detected at mid-infrared wavelengths at the sensitivity level of 
our data.

\item The shape of the spectral energy distribution of TMR-1C is consistent with a
model SED of a very low-mass stellar source of $0.13{\rm M}_{\odot}$ having a circumstellar disk seen almost
edge-on ($i=87^{\circ}$) and located at the distance of the Taurus molecular cloud.
The presence of the edge-on disk that
significantly dims the central source plus additional interstellar extinction of A$_{\rm V}\sim18$\,mag is 
then the cause for the very low apparent brightness of TMR-1C. Hence, the observations at 
near-infrared mainly show TMR-1C in scattered light. 

\item Through our very sensitive H- and Ks-band observations with very good spatial 
resolution we were able to detect a faint point-source that we call TMR-1D (Ks$\sim$18.7\,mag). TMR-1D is located roughly
20$^{\prime\prime}$ north-west of TMR-1AB and, in analogy to TMR-1C, is seen at the end
of a diffuse filament-like dust structure which arises from the north-western nebulosity associated with TMR-1. 
There is an apparent symmetry of TMR-1C and 'its filament' with TMR-1D and 
'its filament'.  TMR-1D is a very low luminosity object, if it is physically associated with TMR-1.

\item The candidate low-mass objects TMR-1C and TMR-1D may have been formed in an
outflow triggered collapse of a string of low-mass cores, or from fragmentation in dense tidal filaments
that were generated during an encounter between protostellar disks of TMR-1A and B. In 
these scenarios the filaments are either dust material swept-up by the outflow, or tidal arms
formed from interacting proto-stellar disks.  

\item Our analysis of spectra extracted at several locations along the filament that 
'points' towards TMR-1C shows the presence of strong molecular hydrogen emission due to shock excitation. 
The physical nature of the shocks is well described by slightly curved C-type bow shock models. 

\item The extinction towards the filament is likely increasing from 
the filament's origin, close to the source TMR-1AB, along the filament in the direction of TMR-1C.
This is implied by extinction values derived from the Q-branch of molecular 
hydrogen at $2.4-2.5\mu$m and means the filament 
structure is probably receding into the denser parts of the molecular cloud.
TMR-1C itself, however, may be sitting slighty in front of the filamant's tip, because
there is no clear indication that the extinction along the line of sight to TMR-1C is higher than
towards TMR-1AB.
We suggest the filament is best interpreted 
in terms of pre-existing dense material being hit by the current outflow, which causes shock
emission. An additional component of the filament is continuum emission from scattered light. 
\end{itemize}
Our results provide new arguments in favour of TMR-1C being a young low-mass object 
associated with the TMR-1 proto-stellar system rather than being
an extincted background star. Moreover, we argued that the faint object TMR-1D may
also be a component of the TMR-1 system. Still, unambiguous proof needs to be supplied.
A confirmation of an edge-on circumstellar disk surrounding TMR-1C would be the detection of
an infrared excess at $\lambda > 10\mu$m and polarized scattered light emission
from a disk and/or an envelope.
The expected flux from a disk at around 20$\mu$m is at a level that is detectable with currently available
mid-IR instrumentation on 8-10\,m class telescopes, and follow-up observations are crucially demanded.
An associated bipolar scattered light structure, however, is probably much more difficult to observe,
since the potential size of the disk (and probably also envelope) is small and is estimated to 
be $< 0.1\arcsec$, based on the point-source morphology at all HST and ISAAC images. 
The nature of TMR-1D is currently undetermined and 
follow-up observations are also needed for this object to fully clarify its nature.

There are currently only very few proto-stellar systems known to harbor very low-luminosity
wide companions in their immediate surroundings. 
Such systems are especially important for probing very low-mass star formation
models, and TMR-1 is potentially one of those prime-target proto-stellar systems.





\begin{acknowledgements}
This work has been carried out based on observations obtained at ESO's LaSilla/Paranal Observatory. 
The authors thank  F. Comeron and F. Bertoldi for valuable suggestions and discussions during 
early stages of this study. We also thank the referee, J.M. Alcal\'a, for his detailed and prompt referee
report which improved the presentation of this paper.
This publication also makes use of data products from the Two Micron All Sky Survey, 
which is a joint project of the University of Massachusetts and the Infrared 
Processing and Analysis Center/California Institute of Technology, funded by 
the National Aeronautics and Space Administration and the National Science 
Foundation.
\end{acknowledgements}
	
{}

\end{document}